\documentclass[10pt,a4paper,twocolumn,showpacs,amsmath,pra,amssymb,english]{revtex4}
\usepackage[T1]{fontenc}
\usepackage[latin9]{inputenc}
\usepackage{float}
\usepackage{graphicx}
\usepackage{amssymb}
\usepackage{esint}
\usepackage{times}

\makeatletter

\usepackage{babel}
\makeatother

\begin{document}

\title{Quantitative test of mean-field description of a trapped two-dimensional Bose gas}

\author{R.~N.~Bisset and P.~B.~Blakie}
\affiliation{Department of Physics, Jack Dodd Centre for Quantum Technology, University of Otago, P.O. Box 56, Dunedin, 9016 New Zealand}


\date{\today}
\begin{abstract}
We investigate the accuracy of two mean-field theories of the trapped two-dimensional Bose gas at predicting transition region properties by comparison to non-perturbative classical field calculations. To make these comparisons we examine the  density profiles and the predictions for the Berezinskii-Kosterlitz-Thouless superfluid transition temperature over a parameter range in which the degree of thermal activation in the tightly trapped direction varies considerably. These results present an important test of these mean-field theories, and provide a characterization of their typical accuracy.
\end{abstract}

\maketitle
\section{Introduction} 
Evidence for the Berezinskii-Kosterlitz-Thouless (BKT) superfluid transition in a trapped two-dimensional (2D) Bose gas  has been reported in a series of experiments by the ENS, NIST and Harvard groups  \cite{Stock2005,Hadzibabic2006,Kruger2007,Clade2009,Gillen2009}. 
Comprehensive theoretical descriptions of the trapped system have been provided by classical field (c-field) and quantum Monte Carlo (QMC) methods \cite{Simula2006,Bisset2009,Bisset2009B,Holzmann2008}, building on the earlier theoretical work for the homogeneous Bose gas \cite{Prokofev2001,Prokofev2002,Posazhennikova2006a}.

It is desirable to have a simple theoretical description for the 2D trapped Bose gas that can be more easily employed than c-field or QMC methods.
In 2008, Holzmann, Chevallier and Krauth (HCK) \cite{Holzmann2008A} presented an important step towards this goal: a semiclassical mean-field theory they used to predict the BKT superfluid transition temperature, $T_{BKT}$.
This work differed from earlier mean-field work (e.g.~see \cite{Bhaduri2000,Gies2004}) which aimed to understand the low temperature properties of the gas, and stability of the Bose condensed phase against thermal fluctuations. 
The HCK theory instead was applied to the high temperature regime (without condensate) to calculate the system density profile. From this the BKT superfluid transition  could be identified by locating the temperature at which the peak phase space density of the gas obtained the critical value determined from studies of the uniform system \cite{Prokofev2001}.
In \cite{Holzmann2008A} several comparisons were made between the HCK theory and full QMC calculations showing good agreement for the density profiles and a prediction of $T_{BKT}$.

In previous work \cite{Bisset2009A} we developed a more complete semiclassical mean-field theory and a consistent procedure identifying  $T_{BKT}$.  For clarity we refer to this  as the full semiclassical (FSC) mean-field theory. While the FSC theory reduces to the HCK with some ad hoc  simplifications, in the degenerate regime the predictions of the two theories differ appreciably. Indeed, the direct comparisons between the two mean-field theories made in  \cite{Bisset2009A}  demonstrated that FSC always predicts lower values for both the system density profile and $T_{BKT}$. Notably, for large systems the differences between the predictions was  significant.
It was argued that the good agreement between the HCK mean-field theory and the QMC calculations was likely fortuitous.  

Due to the increase in interest in the 2D trapped Bose gas it is necessary to better understand the reliability of these mean-field theories for making accurate predictions.
In this work we provide quantitative tests of both mean-field theories against c-field calculations. We present results spanning a broad regime in which the system crosses from being pure-2D (negligible thermal activation in the tightly trapped direction) to quasi-2D (with appreciable thermal activation).
We compare the theories through their predictions for the density profiles and the superfluid transition temperature. Our results provide a benchmark test for the use of mean-field theory, and caution against its use as a quantitative tool. Indeed, our results show that the good agreement observed between HCK mean-field theory and QMC in Ref.~\cite{Holzmann2008A} was due to the quasi-2D regime considered there, and such good agreement is not found in the pure-2D regime.


\section{Quasi-2D mean-field theory}
In this section we review the mean-field theory developed in Refs.~\cite{Holzmann2008A,Bisset2009A}.
We consider a system of ultra-cold bosons confined in a harmonic potential with trap frequencies $\{\omega_x,\omega_y,\omega_z\}$, interacting via a short range potential characterized by the $s$-wave scattering length $a$.
The 2D regime is realized when the trapping potential is sufficiently tight in one direction (which we take to be $z$) that $\hbar\omega_{x,y}\ll k_BT\sim\hbar\omega_z$, with $T$ the system temperature. We limit our consideration to the regime where the scattering is three-dimensional (i.e.~energy independent), which requires that the $z$-confinement length, $a_z = \sqrt{\hbar/m\omega_z}$, satisfies $a_z \gg a$ \cite{Petrov2000}. This requirements is well-satisfied by experiments \cite{Stock2005,Hadzibabic2006,Kruger2007,Clade2009}. It is convenient to define two addition 2D regimes, which we will often refer to: The pure-2D regime, when $k_BT<\hbar\omega_z$, and the quasi-2D regime, when $k_BT\gtrsim\hbar\omega_z$.

\subsubsection{FSC theory}
If interactions are small compared to $\hbar\omega_z$ then the Hartree-Fock modes are of the separable form $\psi_{k\sigma}(x,y,z)=f_{k\sigma}(x,y)\xi_k(z)$, where the axial modes $\xi_k(z)$ are bare harmonic oscillator states.
In the regime of interest the radial plane, for which we introduce the notation $\mathbf{r}=(x,y)$, can be treated semiclassically, eliminating the need to diagonalize for the modes $f_{k\sigma}(\mathbf{r})$ (also see \cite{Hadzibabic2008}). However, the axial modes must be treated quantum mechanically, and the Hartree-Fock expression for the areal density of the system in the $j$-th axial mode is
\begin{equation}\label{nj}
n_j(\mathbf{r}) = \frac{1}{(2\pi)^2}\int d^2\mathbf{k}_r\, \frac{1}{\exp{\left\{\frac{\epsilon_j(\mathbf{r},\mathbf{k}_r)-\mu}{k_BT}\right\}} - 1},
\end{equation}
where  
\begin{eqnarray}
\epsilon_j(\mathbf{r},\mathbf{k}_r) &=& \frac{\hbar^2k_r^2}{2m} +V_j(\mathbf{r}),\label{ej}\\
V_j(\mathbf{r})&=&\sum_{l=x,y}\frac{m}{2}\omega_l^2x_l^2 +j\hbar\omega_z+2\sum^\infty_{k=0}g_{kj}n_k(\mathbf{r}),\label{Vj}
\end{eqnarray}
$\mu$ is the chemical potential, and
\begin{equation}
g_{ki} =\frac{4\pi a\hbar^2}{m}\int dz \,|\xi_k(z)|^2|\xi_i(z)|^2,
\end{equation} describes the interactions between atoms in the $k$ and $i$ axial modes.
Performing the momentum integration in Eq.~(\ref{nj}) and adding up the axial mode densities gives the total areal density
\begin{equation}\label{localden}
n(\mathbf{r}) = -\frac{1}{\lambda_{dB}^2}\sum_{j=0}^\infty\ln \left[1-\exp\ \left(\{\mu-V_j(\mathbf{r})\}/k_BT\right)  \right],
\end{equation}
where
 $\lambda_{dB} = h/\sqrt{2\pi mk_BT}$ is the thermal de Broglie wavelength.
 We refer to the self-consistent solution of Eqs.~(\ref{nj}) and (\ref{ej})  as the full semi-classical (FSC) mean-field theory, to distinguish it from the simplified HCK theory presented below. 
 
\subsubsection{HCK theory}
The theory of HCK \cite{Holzmann2008A} is a simplification of the FSC mean-field scheme presented above made by taking the interactions to be axial mode independent,
 i.e.,  simplifying the mean-field interaction term to
$2\sum_{k=0}^\infty g_{kj}n_k(\mathbf{r})\rightarrow
2g_{\rm{H}}n(\mathbf{r}),$
with the \emph{average} interaction strength 
$g_{\rm{H}}=
{(4\pi a \hbar^2}/{m})\int dz  \rho (z)^2,$
where $\rho(z)$ is the density of a single atom in a harmonic oscillator of frequency $\omega_z$ at temperature $T$.

\subsubsection{Comparison of theories and $T_{BKT}$}\label{TBKT}
An in-depth comparison the FSC and HCK mean-field theories is given in \cite{Bisset2009A}. Most importantly, that study showed that the HCK theory predicts  appreciably different density distributions in the degenerate regime, typically  with higher (peak) densities. 

Mean-field theory can be used to estimate the superfluid transition temperature for the trapped system by using a local density extension of the transition criterion for the uniform system \cite{Prokofev2001}. The transition temperature ($T_{BKT}$) is identified as where the central (peak) phase space density of the system satisfies the condition
\begin{equation}
n_{cr}\lambda_{dB}^2=\ln\left(\frac{\hbar^2C}{m {g}}\right),\label{Eq:peakdencond}
\end{equation}
with $C=380\pm3$  and ${g}$ is the  2D coupling constant.
Within the FSC theory the correct identifications for using Eq.~(\ref{Eq:peakdencond}) is $g=g_{00}$ and $n_{cr}=n_0(\mathbf{0})$  (i.e.~the ground axial mode interaction parameter and density, see discussion in \cite{Bisset2009A}), however HCK proposed using $g=g_{\rm{H}}$ and $n_{cr}=\sum_jn_j(\mathbf{0})$   (i.e.~total areal density).

In combination, the differences in formulation of the two theories and their respective procedures for identifying $T_{BKT}$, is that the critical temperature predicted by FSC ($T_{BKT}^{FCS}$) is lower than HCK value ($T_{BKT}^{HCK}$).


\section{Results}\label{Sec:Results}

To investigate these mean-field theories in the vicinity of the transition we compare their predictions to those obtained by a c-field method.
The c-field method is based on a classical field representation of the highly occupied low energy modes (i.e.~those below an appropriately chosen energy cutoff, see \cite{Blakie2007}) and a Hartree-Fock treatment of the sparsely occupied (high energy) modes of the system. This method treats the highly occupied (strongly fluctuating)  modes non-perturbatively and is valid in the critical region. For details of this c-field theory refer to \cite{Blakie2008} and for the specific application to the quasi-2D trapped Bose gas see \cite{Simula2006,Simula2008,Bisset2009}.
All results we present here are for $^{87}\rm{Rb}$ atoms in cylindrically symmetric trap with radial trapping frequencies $\omega_x=\omega_y=2\pi\times9.4$ Hz and axial frequency of either $\omega_z=2\pi\times0.94$ kHz or $\omega_z=2\pi\times1.88$ kHz.

\subsection{System density in the transition region}\label{Sec:sysden}
\begin{figure}
\centering
\includegraphics[width=3.0in]{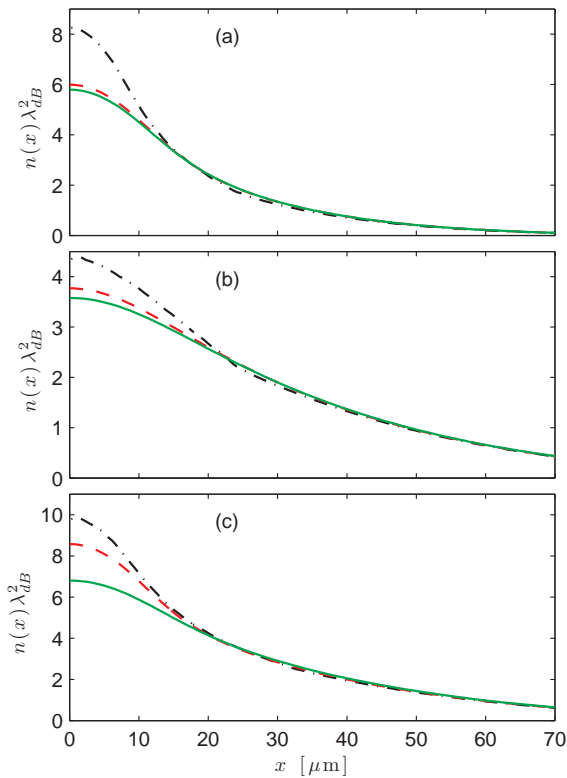}
\caption{\label{Fig:MeanF_CFS_den_ca} Phase space density. Parameters:  (a) , $N = 13.4\times10^3$, $T=34.7$ nK ($\approx T_{BKT}$). (b)  $N=45.8\times10^3$, $T=67.6$ nK. (c) $N=67.4\times10^3$, $T=65.0$ nK ($\approx T_{BKT}$). Curves: FSC  (solid),  HCK  (dashed), and c-field   (dot-dashed) results. Other parameters:  $\omega_{x,y}/2\pi=9.4$ Hz for all results, $\omega_z/2\pi=1880$ Hz in (a) and (b), and $\omega_z/2\pi=940$ Hz in (c).}
\end{figure}

An important quantity for detailed comparison is the system density profiles. In the development of the HCK theory direct comparisons of the density profiles were made to the results of quantum Monte Carlo calculations \cite{Holzmann2008A}, however results were only given  for cases where the peak phase space density was less than half the critical value (\ref{Eq:peakdencond}), i.e.~in relatively non-degenerate regime where the mean-field description is expected to work well. Here we present results from regimes closer to the transition.

Figure \ref{Fig:MeanF_CFS_den_ca} shows density profile predictions of the various theories at several temperatures. 
The case shown in Fig.~\ref{Fig:MeanF_CFS_den_ca}(a) is approximately at the transition temperature (as determined by the c-field method \cite{Bisset2009}), and reveals a rather large difference between the c-field and mean-field predictions. In this regime both mean-field theories fail because they do not describe the quasi-condensate (suppression of density fluctuations) that forms \cite{Safonov1998}. Quasi-condensation allows the system density to greatly increase, explaining why the c-field density exceeds the mean-field density. 
The onset of quasi-condensation is not abrupt at the transition, but gradually forms as a precursor to the transition at temperatures well above  $T_{BKT}$ \cite{Prokofev2002} (e.g.~in Fig.~\ref{Fig:MeanF_CFS_den_ca}(b) where a small quasi condensate component [about 3.6\% of the system, but quite dominate near trap center] causes the c-field density to be slightly higher than the mean-field predictions near $\mathbf{r}=\mathbf{0}$). The results in Fig.~\ref{Fig:MeanF_CFS_den_ca}(a) and (b) are in the pure-2D regime, and in Fig.~\ref{Fig:MeanF_CFS_den_ca}(c) we consider a case near $T_{BKT}$, but in the quasi-2D regime. For this case, the appreciable thermal activation in the $z$-direction reduces $g_{\rm{H}}$ from $g_{00}$ and the HCK theory is closer to the c-field result.

\subsection{Comparison of predictions for $T_{BKT}$}\label{Res:extrapo}

\begin{figure}
\centering
\includegraphics[width=3.0in]{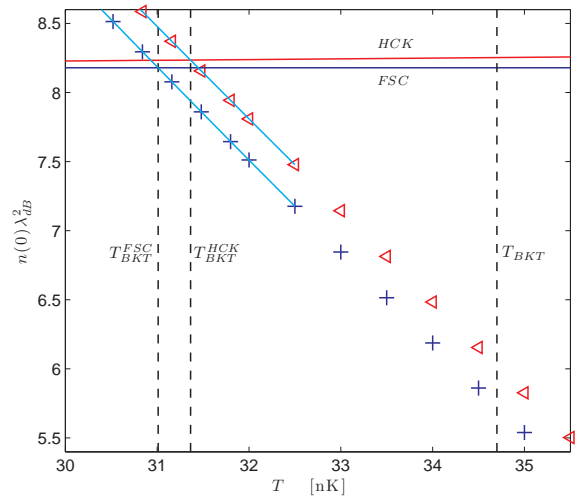}
\caption{\label{Fig:MeanFBKT_N11267A} Peak phase space density as a function of $T$.  FSC value of $n_0(\mathbf{0})\lambda_{dB}^2$ (pluses), HCK peak value of the total areal phase space density, $\sum_jn_j(\mathbf{0})\lambda_{dB}^2$ (triangles) and polynomial fits to these results in the transition region are shown as lines. Also displayed (as nearly horizontal solid lines) are the critical phase space densities for the FSC and HCK theories (see Eq.~(\ref{Eq:peakdencond})). The predictions for the transition temperature made by the FSC ($T_{BKT}^{FSC}$), the HCK  ($T_{BKT}^{HCK}$) and  c-field  ($T_{BKT}$) theories are indicated as vertical dashed lines. Calculation parameters: $\omega_{x,y}=2\pi\times9.4$ Hz, $\omega_{z},=2\pi\times1.88$ kHz and $N=13.5\times10^3$.}
\end{figure}

We can now compare the predictions of the FSC and HCK mean-field theories for the transition temperature $T_{BKT}$ against the c-field simulations. 
In Fig.~\ref{Fig:MeanFBKT_N11267A} we present an example of how this comparison is made. For this case the system has a large aspect ratio (i.e.~$\omega_z/\omega_{x,y}=200$) and in the transition region the system is in the {pure 2D} limit where $g_{00}\approx g_{\rm{H}}$, and the difference between $n_0(\mathbf{r})$ and the total areal density is negligible.

For the two mean-field theories we calculate the peak phase space density at each temperature and then locate the transition temperature by condition (\ref{Eq:peakdencond}) appropriately interpreted for each theory (see discussion in Sec.~\ref{TBKT}).
The respective critical phase space density for each theory is shown Fig.~\ref{Fig:MeanFBKT_N11267A},
 however for the pure-2D regime of this calculation there is little difference between these values, and we find that $T_{BKT}^{FSC}$ and $T_{BKT}^{HCK}$ are similar to each other but are both significantly reduced from the c-field prediction for $T_{BKT}$.

\begin{table}
\centering
\begin{tabular}{ c | c |c |c |c |c}
\hline
\hline
& \multicolumn{5}{c}{Cases}\\
\hline
& A & B & C & D & E \\
\hline
\hline
$N/10^3$ & 13.5 & 21.2 & 30.2 & 67.3  & 228.5 \\
$\omega_z/2\pi$ & 1880 Hz & 1880 Hz & 940 Hz & 940 Hz  & 940 Hz\\
$T_{BKT}$ & 34.7 nK & 43.1 nK & 46.8 nK & 64.9 nK & 104 nK  \\
$\nu_z$ & 0.38 & 0.48 & 1.04 & 1.44  & 2.32\\
\hline
$T^{FSC}_{BKT}$ & 31.0 nK & 38.4 nK & 43.5 nK & 60.4 nK  & 96.9 nK  \\
rel.~err. & -0.11 & -0.11 &   -0.07 &  -0.07  & -0.07\\
$\ln\left(\frac{380\hbar^2}{mg_{00}}\right)$ & 8.18 & 8.18 &   8.52 & 8.52  & 8.52\\

\hline
$T^{HCK}_{BKT}$ & 31.4 nK & 39.2 nK & 45.8 nK & 64.4 nK  & 105 nK  \\
rel.~err.  & -0.10 & -0.09 & -0.02 &    -0.01   & +0.01\\ 
$\ln\left(\frac{380\hbar^2}{mg_{\rm{H}}}\right)$ & 8.23 & 8.28 &  8.92 & 9.07  & 9.30\\
\hline
\hline
\end{tabular}
\caption{\label{Tab:trap_int} Comparison of mean-field predictions for the critical temperature. The c-field prediction for the critical temperature ($T_{BKT}$) is used to obtain $\nu_z$. The relative error of each mean-field estimate of $T_{BKT}$ is given against the c-field value and the radial trap frequencies in all cases are $\omega_{x,y}/2\pi=9.4$ Hz. Interaction strength $g_H$ calculated using corresponding temperature $T_{BKT}^{HCK}$. Logarithms indicate critical phase space density condition for each theory.}
\end{table}

In Table \ref{Tab:trap_int} we repeat the analysis shown in Fig.~\ref{Fig:MeanFBKT_N11267A}  for a system in five different parameter regimes, which we refer to as cases A-E. Importantly, these results explore parameters whereby the system crosses over from being pure-2D to quasi-2D at the transition point, as quantified by the parameter $\nu_z=k_BT_{BKT}/\hbar\omega_z$. In the pure-2D regime ($\nu_z<1$, cases A, B) we find that the FSC and HCK theories make similar predictions, and both significantly underestimate the transition temperature. This occurs because they fail to account for the suppression of  density fluctuations and thus underestimate the phase space density of the system, as observed in Fig.~\ref{Fig:MeanF_CFS_den_ca}(a).
 In the quasi-2D regime ($\nu_z \gtrsim 1$, cases C, D, E) the HCK prediction agrees quite well with the c-field result for $T_{BKT}$, while the FSC remains an (appreciable) underestimate. The main origin of the improved agreement is the reduction in $g_{\rm{H}}$ that occurs as $\nu_z$ increases, which tends to increase the HCK density. We emphasize this increase is not for the correct physical reason  (suppressed density fluctuations), but simply from the ad hoc choice of $g_{\rm{H}}$. Figure \ref{Fig:MeanF_CFS_den_ca} (c) shows the density profile near $T_{BKT}$ for case D. We also note that 
 $T_{BKT}$ is identified in the c-field simulations using the same criterion used for the FSC theory, i.e.~when the peak phase space density of the ground axial mode satisfies the condition  (\ref{Eq:peakdencond}) with $g=g_{00}$.

In Ref.~\cite{Holzmann2008A} the comparison of the HCK prediction for  $T_{BKT}$ with QMC is made for a system with $\nu_z\approx1.4$, and agreement is found to be at the few percent level. In the context of our results here, it is clear that this good agreement is likely facilitated by the reduction in $g_{\rm{H}}$ from $g_{00}$ compensating for the absence of quasi-condensate in the mean-field theory.

\section{Conclusions}

We have compared the density predictions made by two semiclassical mean-field theories against the c-field method.  At low temperatures, we find that both mean-field theories overestimate interaction effects due to their neglect of correlations that cause density fluctuation suppression (quasi-condensation). In this regime both methods systematically underestimate the central density, and underestimate $T_{BKT}$. 

For larger systems, where the transition occurs at temperature with appreciable thermal excitation in the tightly trapped mode, the scenario is quite different. 
The HCK approximation of replacing the mode dependent interaction parameter by a thermally averaged value, leads to an increase in the system density, and a higher prediction for the transition temperature than the FSC theory. 
We expect that for sufficiently large systems the HCK theory will over estimate $T_{BKT}$, although we have not presented results here that show this.

The quantitative tests we have presented provide a useful characterization of the applicability of mean-field methods in the vicinity of the BKT transition for the 2D trapped Bose gas. Our results are over a broader range of parameters than those considered in \cite{Holzmann2008A}, and thus present a more balanced view of the method effectiveness.
 Generally, our results show that mean-field methods cannot be relied upon for accurate predictions, but do nevertheless  provide  a rough estimate of the degeneracy regime. With future experiments likely producing tighter 2D traps, in which degeneracy will occur in the pure-2D regime, our observations on the failure of mean-field theory in this regime will be important. Of course, in the quest for a simple and effective mean-field theory several other factors must be borne in mind: (i) The HCK theory involves a single interaction parameter allowing it to be computed extremely efficiently using the procedure presented in \cite{Holzmann2008A}. (ii) The FSC theory provides a lower bound for $T_{BKT}$.

{\bf Acknowledgments} PBB and RNB are supported by NZ-FRST contract NERF-UOOX0703. RNB acknowledges support from the Otago Research Committee.

\end{document}